\documentclass[conference]{IEEEtran}

\pdfoutput=1

\newcommand{\RR}{\mathbb{R}}

\newcommand{\NN}{\mathbb{N}}

\newcommand{\eu}  {\mathrm{e}}
\newcommand{\jj}  {\mathrm{j}}

\newcommand{\snr} {\mathsf{SNR}}

\newcommand{\E}{\mathbb{E}}
\newcommand{\reff}[1]{\eqref{#1}}
\newcommand{\defeq}{\triangleq}

\newcommand{\cv}{\boldsymbol{c}}

\newcommand{\rv}{\boldsymbol{r}}

\newcommand{\Xv}{\boldsymbol{X}}

\newcommand{\Hm}{\boldsymbol{H}}
\newcommand{\Id}{\boldsymbol{I}}

\newcommand{\Nc}{\mathcal{N}}

\newcommand{\Xc}{\mathcal{X}}

\newcommand{\rhov}{\boldsymbol{\rho}}

\newcommand{\Sigmam}{\boldsymbol{\Sigma}}

\usepackage{cite}      

\usepackage{graphicx}  

\usepackage{psfrag}    

\usepackage{subfigure} 

\usepackage{url}       

\usepackage{stfloats}  

\usepackage{amsmath}   
\usepackage{array}
 \makeatletter
 \let\NAT@parse\undefined
 \makeatother
\usepackage{times}
\usepackage{epsfig,graphicx,color,pstricks}
\usepackage{amsmath,amssymb,amsbsy,amsthm,amsfonts,latexsym}
\usepackage{bm}
\usepackage{psfrag}
\usepackage[mathscr]{eucal}
\usepackage{cite}
\usepackage[all]{xy}
\usepackage{subfigure}
\usepackage{graphics}

\newtheorem{theorem}{Theorem}
\newtheorem{lemma}[theorem]{Lemma}

\newtheorem{remark}{Remark} 

\begin{document}

\title{A New Sum-Rate Outer Bound for Interference Channels with Three Source-Destination Pairs}

\author{%
\authorblockN{Daniela Tuninetti}
\authorblockA{Department of Electrical and Computer Engineering,\\
University of Illinois at Chicago, Illinois 60607, USA,\\
Email: danielat@uic.edu}
}
\maketitle

\begin{abstract}
This paper derives a novel sum-rate outer bound for the general memoryless
interference channel with three users.  The derivation
is a generalization of the techniques developed by Kramer and 
by Etkin {\em et al} for the Gaussian two-user channel. 
For the Gaussian channel the proposed sum-rate outer bound
outperforms known bounds for certain channel parameters.
\end{abstract}


\section{Introduction}
\label{sec:intro}
An interference channel models an ad-hoc wireless network where
several uncoordinated source-destination pairs share the same 
channel thereby creating undesired mutual interference at the
receivers.  Today's networks
are designed to avoid interference through resource division among
users because interference is considered the bottleneck
of high-speed data networks. It is well known however that user
orthogonalization, in frequency, time, space or code domain,
is in general suboptimal in terms of performance.
With advances in computing technology,
it has become possible to design communication strategies
to {\em manage} the interference.
This trend has renewed the interest in the ultimate limits
of interference networks. Much progress has been made in the past
few years on understanding the capacity of the Gaussian interference
channel with two source-destination pairs.
However, interference channels with more than two source-destination pairs,
or non-Gaussian channels, are far less understood.
The objective of this work is to investigate the maximum throughput,
or sum-rate, or sum-capacity,
of the general memoryless interference channel with three source-destination pairs.
The generalization of the proposed bounding technique 
to the whole capacity region and to an arbitrary number of source-destination pairs
is presented in~\cite{tuninetti-outer:KIFC:ISIT2011}.

Before revising past work on interference networks and outlining
our main contributions, we formally introduce
the network problem considered in this paper.

\subsection{Problem Definition}
\label{subsect:genchmodel}

\begin{figure}
	\centering
		\includegraphics[width=6cm]{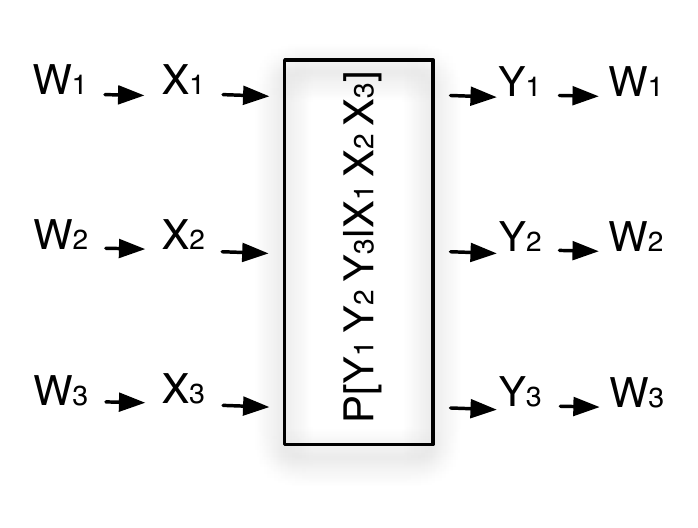}
	\vspace*{-0.5cm}
	\caption{The general memoryless InterFerence channel with three source-destination pairs (3-IFC) considered in this work.}
	\label{fig:DMCchannel}
\end{figure}

Our notation follows the convention in~\cite{LNIT}.
The channel considered in this work is depicted in Fig.~\ref{fig:DMCchannel}.
An InterFerence Channel with three source-destination pairs
(3-IFC) is a multi-terminal network where source $i$, $i\in\{1,2,3\}$,
wishes to communicate to destination $i$ through a shared memoryless
channel with transition probability $P_{Y_1,Y_2,Y_3|X_1,X_2,X_3}$.
Each source $i$, $i\in\{1,2,3\}$, encodes an independent message $W_i$ 
of rate $R_i\in\RR_+$ into a codeword of length $n\in\NN$.
We adopt standard definitions of codes, achievable rates and capacity region~\cite{LNIT},
that is, the capacity region is the convex closure
of the set of rate-triplet $(R_1,R_2,R_3)$ 
for which the error probability 
goes to zero as the block-length $n\to \infty$.
As for other channels without destination cooperation, 
the capacity region of the 3-IFC only depends 
on the channel marginals $P_{Y_k|X_1,X_2,X_3}$, $k\in\{1,2,3\}$,
and not on the whole joint channel transition probability $P_{Y_1,Y_2,Y_3|X_1,X_2,X_3}$.

\subsection{Past Work}
\label{subsec:past}
The capacity region of a general memoryless 3-IFC is not known
-- not even the capacity of the 2-IFC has been characterized
in full generality at present.

For a 2-IFC, the capacity region is known if:
a1) the interfering signal is strong at each destination~\cite{sato:IFCoutit1978,costa_elgamal:strong:it1987}; if the interference is very strong, then interference does not reduce capacity~\cite{carleial:ifcstrong:it1975};
b1) the outputs are deterministic functions of the inputs and invertible 
given the intended signal~\cite{costa_aelgamal:it1982}; and
c1) the channel has a special form of degradeness~\cite{sato:IFCoutit1978,liu_ulukus:degifc,benzel:it1979 }.
The largest known achievable region, due to Han and Kobayashi (HK)~\cite{Han_Kobayashi:it1981}, uses
rate-splitting and simultaneous decoding of the intended message and part of the interfering message.
The best outer bound without auxiliary random variables is due to Sato~\cite{sato:IFCoutit1977},
and with auxiliary random variables is due to Carleial~\cite{carleial:IFCit1983} (see also Kramer~\cite[Th.5]{kramer:it2004}).

For the Gaussian 2-IFC, the capacity region is fully known
in strong interference only~\cite{carleial:ifcstrong:it1975,sato:IFCstrong:it1981,costa:awgnifc:it1985}, that is, when
the interference is strong at each destination.
The sum-capacity is however known in the following cases: 
a2) in mixed interference~\cite{tuninettiweng:isit2008,canadanoisy}, that is, when
one interfering signal is strong  and the other is weak;
b2) for the Z-channel~\cite{sason:it2004}, that is, when only one destination
experiences interference; and
c2) in very weak interference at both destinations~\cite{biaonoisy2008:IT09,venunoisy2008:IT09,canadanoisy}.
In the mixed and weak interference regimes, a simple message-splitting in the HK region 
is to within two~bits~\cite{caire2bits} of the outer bound proposed
in~\cite{etkin_tse_hua:withinonebit:subIt06} for all channel parameters.
The best outer bound for the  Gaussian 2-IFC is obtain by intersecting the regions
derived by Kramer in~\cite[Th.1]{kramer:it2004} and in~\cite[Th.2]{kramer:it2004},
by Etkin {\em et al.} in~\cite{etkin_tse_hua:withinonebit:subIt06}, and the region independently obtained
in~\cite{biaonoisy2008:IT09,venunoisy2008:IT09,canadanoisy} and later further tighten by Etkin in~\cite{EtkinISIT2009}.

Very few results are available for a general memoryless K-IFC with $K\geq 3$.
General inner bound regions are lacking.
A straightforward generalization of the HK approach, 
whereby each user has a different (sub)message for every subset of 
non-intended receivers, would require the specification of
$K 2^{K-1}$ (sub)rates. The resulting region
would have $K 2^{(K+1)2^{K-2}-1}$ bounds and
would still require an application of the
Fourier-Motzkin elimination procedure in order to be expressed
as a function of $K$ rates only.
Thus the HK approach for more than two users
appears impractical because of its super-exponential complexity
in the number of users.
The HK approach might also be suboptimal in general.
In fact, decoding at each receiver in a K-IFC is impaired by the joint effect of
all the interferers, rather by each interferer separately.
Consequently, coding schemes that deal directly with
the effect of the combined interference could have superior performance
in terms of achievable rates than the HK approach.  Examples of such coding
schemes for the Gaussian K-IFC are {\em interference alignment}~\cite{Cadambe-Jafar:arXiv:0707.0323v2}
and {\em structured codes}~\cite{Nazer:2007, Bresler:1toM_Mto1:allerton07, Sridharan:layredlattice:allerton08}.

In Gaussian noise, channels with special structure have been investigated:
a3) the ``fully symmetric'' 3-IFC, whereby all interfering links
have the same strength and all direct links have the same strength,
was considered in~\cite{cadambe:notseparable:it09}; a genie-aided outer
bound that provides a group of receivers with sufficient side information
so that they can decode a subset of the users as in a Multiple Access Channel (MAC) channel
was also discussed in~\cite{cadambe:notseparable:it09} and was later generalized in~\cite{josesriram:kifcmacbound:itw10} to any number of users and any general channel matrix structure (however the resulting outer bound appears very difficult to evaluate in closed form); 
b3) the ``cyclic symmetric'' channel, whereby all receivers have a statistically
equivalent output up to cyclic permutation of the user indices and are interfered by one
other user only, was considered in~\cite{ZhouYu:cyclicsym:CISS10}; it was shown that a generalization
of the approach of~\cite[Th.1]{etkin_tse_hua:withinonebit:subIt06} 
gives capacity to within two~bits when the interference is weak;
if instead the interference is strong, the whole capacity region is given by an application of~\cite[Th.1]{kramer:it2004} to each receiver;
c3) the high-SNR linear deterministic approximation of the ``cyclic symmetric'' 3-IFC,
without the restriction of having one-sided interference as in~\cite{ZhouYu:cyclicsym:CISS10},
was studied in~\cite{Bandemer:cyclicsym:isit09}; the sum-capacity was characterized for almost all
choices of parameters for the case where one interferer is strong and the other 
is weak;  the corresponding finite-SNR model was not discussed;
d3) the ``cyclic mixed strong-very strong'' 3-IFC was studied in~\cite{segzin_3ifc:arXiv:1010.4911};
here again the whole capacity region is obtained by applying~\cite[Th.1]{kramer:it2004} to each receiver, assuming that each receiver $k\in\{1,2,3\}$ experiences strong interference from
user $k-1$ and very strong interference from user $k+1$ (indices are defined modulus $3$);
the conditions given in~\cite{segzin_3ifc:arXiv:1010.4911} for the achievability of the outer bound are sufficient;
e3) the one-to-many (only one source creates interference)
and the many-to-one (only one destination experiences interference)
channels were studied in~\cite{Bresler:1toM_Mto1:allerton07}; in both cases capacity
was determined up to a constant number of bits that is an increasing function of
the number of users; the central contribution is to show that
purely random codes (according to the definition in~\cite{Nazer:2007}),
like in the HK scheme, fail to achieve the outer bound to within a constant gap;
instead structured lattice codes are necessary to establish capacity to within a 
finite number of bits; as mentioned before, structured codes are well
suited for multi-interferer problems because they deal with the
aggregate interference seen at a destination; in particular, with
lattice codes, each destination has to decode one ``virtual'' interferer
no matter how many users are present in the network;
f3) continuing on the advantages of structured codes, it is known that
the notion of strong interference does not extend to $K\geq 3$
users in a straightforward manner~\cite{Sridharan:verystrongKifc:globecom08}
and that structured codes outperform purely random codes;
in particular, lattices allow for an ``alignment'' of the interference observed
at each receiver and can achieve the interference-free capacity under 
a milder requirement on the channel matrix than random codes~\cite{Sridharan:verystrongKifc:globecom08};
finally,
g3) the Degrees of Freedom (DoF) of the K-IFC
was considered in~\cite{jafar:dofKifc:it10,EtkinOrdentlichISIT2009,Sridharan:layredlattice:allerton08} and references therein; 
in general, random codes that generalize the two-layer coding schemes of HK to the K-user case are strictly outperformed by lattice codes~\cite{Sridharan:layredlattice:allerton08};
``interference alignment'' is known to achieve $K/2$ DoF for certain channels~\cite{Cadambe-Jafar:arXiv:0707.0323v2};  it is however known that the DoF is discontinuous at all fully connected, rational gain matrices~\cite{EtkinOrdentlichISIT2009}; this points out that high-SNR analysis in problems with many parameters (like the K-IFC) is very sensitive to the way the different parameters are let grow to infinity; the generalized DoF analysis~\cite{etkin_tse_hua:withinonebit:subIt06} appears more appropriate but its complexity is quadratic in the number of users; the generalized DoF of the fully symmetric K-IFC for any $K\geq 2$ is the same as that the 2-IFC except when all channel outputs are statistically equivalent~\cite{jafar:dofKifc:it10} (in which case time division is optimal).

\subsection{Contributions and Paper Organization}
The central contribution of this paper is to propose a {\em framework
to derive sum-rate outer bounds for the 3-IFC 
that naturally generalizes to the whole capacity region
of any memoryless IFC with an arbitrary number of users}~\cite{tuninetti-outer:KIFC:ISIT2011}.
Our contributions are as follows:
1) In Section~\ref{sec:maindmc} we derive a sum-rate outer bound for the {\em general
memoryless 3-IFC} that generalizes the techniques originally
developed by Kramer~\cite{kramer:it2004} and by Etkin {\em et al}~\cite{etkin_tse_hua:withinonebit:subIt06}
for the Gaussian 2-IFC;
2) In Section~\ref{sec:gauss} we evaluate the bound derived
in~Section~\ref{sec:maindmc} for the Gaussian channel.
We show that {\em the proposed bound
improves on existing bounds for certain channel parameters.}
Section~\ref{sec:conc} concludes the paper.

\section{Main Result for the General 3-IFC}\label{sec:maindmc}

We divide the presentation of our novel sum-rate outer bound
into two parts: Th.\ref{th:out1} 
generalizes the approach of Kramer~\cite[Th.1]{kramer:it2004}
and Th.\ref{th:out2} 
generalizes the approach 
of Etkin  {\em et al}~\cite[Th.1]{etkin_tse_hua:withinonebit:subIt06}.  Our proposed outer bound
in the intersection of the regions in Th.\ref{th:out1} and Th.\ref{th:out2}.

\begin{theorem}
\label{th:out1}
The sum-rate of a general memoryless
3-IFC is upper bounded by:
\begin{align}
  R_{1}+R_{2}+R_{3}
  &\leq 
I( Y_{1};X_{1},X_{2},X_{3},Q)
\nonumber\\&+
I( Y_{2};X_{2},X_{3}| X_{1},Y_{1},Q)
\nonumber\\&+
I( Y_{3};X_{3}| X_{1},Y_{1}, X_{2},Y_{2},Q),
\label{eq:th:out1}
\end{align}
for some input distribution
$P_{X_1,X_2,X_3,Q}=P_{Q}\prod_{k=1}^{3}P_{X_k|Q}$.
By exchanging the role of the users in~\reff{eq:th:out1},
other $(3!-1)=5$ sum-rate bounds can be obtained.
Moreover, 
the sum-rate bound in~\reff{eq:th:out1} can be minimized with respect to 
the joint probability $P_{Y_1,X_3,Y_3|X_1,X_3,X_3}$
as long as the marginal probabilities $P_{Y_k|X_1,X_2,X_3}$, $k\in\{1,2,3\}$,
are preserved.
\end{theorem}
\begin{IEEEproof}
By Fano's inequality:
\begin{align*}
   &n(R_1+R_2+R_3) \leq \sum_{k=1}^{3} I(W_k; Y_k^n) 
\\ &\leq  \sum_{k=1}^{3} I(W_k; Y_k^n, \ Y_{k-1}^n,W_{k-1}, \ldots, Y_{1}^n,W_{1})
\\ &= H(Y_1^n) + H(Y_2^n|Y^n_1,W_1) + H(Y_3^n|Y^n_2,Y^n_1,W_1,W_2)
\\ &\qquad  - H(Y_3^n,Y^n_2,Y^n_1|W_1,W_2,W_3).
\end{align*}
By continuing with standard inequalities (see~\cite{tuninetti-outer:KIFC:ISIT2011} for details)
the bound in~\reff{eq:th:out1} can be obtained.
%
The joint channel transition probability can be optimized so as to tighten the 
sum-rate bound in~\reff{eq:th:out1}, subject to
preserving the marginals, because the capacity region 
only depends on the channel conditional marginal 
probabilities~\cite{sato:out_BC}.
\end{IEEEproof}

\medskip
We remark that:
\begin{enumerate}

\item
The proposed bound reduces to~\cite[Th.1]{kramer:it2004} for the
Gaussian 2-IFC when $\Xc_3=\emptyset$ (see~\cite[eq.(34)]{kramer:it2004}).
Th.\ref{th:out1} however holds for {\em any} memoryless IFC.

\item
As described in~\cite{tuninetti-outer:KIFC:ISIT2011},
Th.\ref{th:out1} can be extended to {\em any number of users} $K$
and to {\em any partial sum-rate}, in which case
the derived region contains $N(K)=\sum_{k=1}^{K}{K \choose k} k!$ bounds.
For $K=2$, the region has $N(2)=4$ bounds as in~\cite[Th.1]{kramer:it2004}
(two single-rate bounds and two sum-rate bounds).
For $K=3$, the region has $N(3)=15$ bounds, of which the ${3 \choose 3} 3!=6$ sum-rate bounds
cannot be derived 
by silencing one of the users and by applying~\cite[Th.1]{kramer:it2004} to the resulting 2-IFC 
(see Section~\ref{sec:gauss}) and are the novel contribution of Th.\ref{th:out1}.

\item
Every mutual information term in
Th.\ref{th:out1} contains {\em all} the inputs $(X_1,X_2,X_3)$
and no auxiliary random variable. This implies that the bound
can be easily evaluated for many channels of interest, including
the Gaussian channel (see Section~\ref{sec:gauss}).

\item
Th.\ref{th:out1} can be easily extended to
memoryless channels without receiver cooperation. 
For example, the 2-IFC with generalized feedback
(a.k.a. source cooperation)~\cite{ifcgf-tuninetti-journal1}
was studied in~\cite{tuniISIT09out,tuniITA10out} and the
extension to any number of users is discussed in~\cite{tuninetti-outer:KIFC:ISIT2011}.
The 2-user cognitive channel was considered in~\cite{RTDjournal1}
and the 2-IFC with a cognitive relay in~\cite{rini2010CIFC+CR:ITWDublin}.

\end{enumerate}

\begin{theorem}
\label{th:out2}
The sum-rate of a general memoryless
3-IFC is upper bounded by:
\begin{align}
&R_1+R_2+R_3
\nonumber\\&
\leq \sum_{k=1}^{3}H(Y_k|S_k,Q)-H(S_k|Y_k,X_1,X_2,X_3,Q),
\label{eq:th:out2}
\end{align}
for some input distribution
$P_{X_1,X_2,X_3,Q}=P_{Q}\prod_{k=1}^{3}P_{X_k|Q}$ and such that 
side information set $\{S_k, k\in\{1,2,3\} \}$ coincides with
the set $\{Y_{\backslash k}, k\in\{1,2,3\} \}$, where
$Y_{\backslash k} \sim Y_k|X_k$, i.e., 
$Y_{\backslash k}$ is statistically
equivalent to the channel output at destination~$k$
from which the intended signal $X_k$ has been removed. 
Moreover,
the sum-rate bound in~\reff{eq:th:out2} can be minimized with respect to
the joint probability $P_{Y_k, S_k|X_1,X_2,X_3}$, $k\in\{1,2,3\}$, as long as
the conditional marginal distributions are preserved.
\end{theorem}
\begin{IEEEproof}
By Fano's inequality:
\begin{align*}
  &n(R_1+R_2+R_3) \leq \sum_{k=1}^{3} I(W_k; Y_k^n)
\leq \sum_{k=1}^{3} I(X_k^n; Y_k^n, S_k^n)
\\&=    \sum_{k=1}^{3} H(S_k^n)
                      -H(Y_k^n|X_k^n)
                      +H(Y_k^n|S_k^n)
                      -H(S_k^n|X_k^n,Y_k^n).
\end{align*}
By assuming that 
$\sum_{k=1}^{3} H(S_k^n)\leq \sum_{k=1}^{3}H(Y_k^n|X_k^n) $ (which is the case
when the side information set $\{S_k, k\in\{1,2,3\} \}$ coincides with
the set $\{Y_{\backslash k}\sim Y_k|X_k, k\in\{1,2,3\} \}$) and
by continuing with standard inequalities (see~\cite{tuninetti-outer:KIFC:ISIT2011} for details)
the bound in~\reff{eq:th:out2} can be obtained. 
\end{IEEEproof}

\medskip
We remark that:
\begin{enumerate}

\item
The proposed bound reduces to~\cite[Th.1]{etkin_tse_hua:withinonebit:subIt06} for the Gaussian 2-IFC
when $\Xc_3=\emptyset$ by setting $S_1 = Y_{\backslash 2}$
and  $S_2 = Y_{\backslash 1}$.
Th.\ref{th:out2} is however tighter than~\cite[Th.1]{etkin_tse_hua:withinonebit:subIt06} for the Gaussian
2-IFC because the correlation between the Gaussian noise of the channel
output $Y_j$ and the Gaussian noise of the side information $Y_{\backslash k}$,
$(j,k)\in\{1,2,3\}^2$, can be optimized (see Section~\ref{sec:gauss}).

\item
Th.\ref{th:out2} holds for {\em any} memoryless 3-IFC. 

\item
Th.\ref{th:out2}
can be extended to {\em any number of users} $K$
and to {\em any partial sum-rate}; some of the bounds in the
so-derived region 
cannot be obtained by simply silencing all but two users and 
then applying~\cite[Th.1]{etkin_tse_hua:withinonebit:subIt06} to the resulying 2-IFC
and are the novel contribution of Th.\ref{th:out2}.

\item
Extensions of Th.\ref{th:out2} to other channel models are possible but 
appear more involved than those of Th.\ref{th:out1}
Such an extension has been presented in~\cite{tuniITA10out,vinodSC2009}
for the 2-IFC with generalized feedback
and in~\cite{rini2010CIFC+CR:ITWDublin} for the 2-IFC with a cognitive relay.

\end{enumerate}

\section{Gaussian channels}\label{sec:gauss}

The Gaussian channel model is introduced in Subsection~\ref{sec:gauss mod}.
A sum-rate outer bound derived form the results available for
the Gaussian 2-IFC is described in Subsection~\ref{sec:gauss known sumrate}.
Subsection~\ref{sec:gauss th1 and th2} evaluates
Th.\ref{th:out1} and Th.\ref{th:out2}.
Subsection~\ref{sec:gauss examples} numerically compares the
proposed sum-rate bounds with some of the results available in the literature
and shows that there are channel parameters for which our
proposed sum-rate bound is the tightest.

\subsection{The Gaussian Channel Model}\label{sec:gauss mod}

\begin{figure}
	\centering
		\includegraphics[width=8cm]{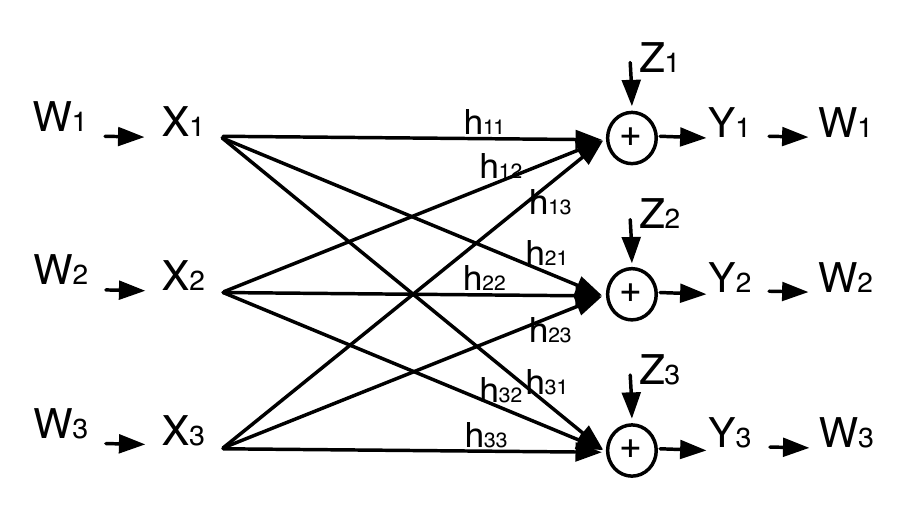}
	\vspace*{-0.5cm}
	\caption{The Gaussian 3-IFC considered in this work.}
	\label{fig:channel}
\end{figure}

A SISO (single input single output) complex-valued
Gaussian 3-IFC in {\em standard} form, depicted in Fig.~\ref{fig:channel},
has outputs:
\begin{align*}
     Y_{i} &= \sum_{k=1}^{3} h_{i,k}X_{k} + Z_{i}, 
\end{align*}
with input power constraint $\E[|X_{i}|^2] \leq 1$ and noise 
$Z_{i} \sim \Nc(0,1)$, $i \in\{1,2,3\}$.
The correlation among the Gaussian noises is irrelevant since the
capacity only depends on the marginal noise distributions.
The channel gains are fixed for the whole transmission duration
and are known to all terminals.
Without loss of generality, the direct link gains $h_{i,i}$, $i \in\{1,2,3\}$,
can be taken to be real-valued (because receiver $i$
can compensate for the phase of $h_{i,i}$)
and strictly positive (if $|h_{i,i}|^2=0$ then the SNR at receiver $i$ is
zero even in absence of interference, which implies that $R_i=0$, i.e., $X_i=0$
is optimal and the system has effectively one less user).
The Gaussian 3-IFC model is completely specified
by the $3\times 3$ channel matrix
$\Hm:  [\Hm]_{i,j}=h_{i,j}$, $(i,j) \in\{1,2,3\}\times\{1,2,3\}$.


\subsection{Known sum-rate bounds}\label{sec:gauss known sumrate}
Sum-rate bounds for the 3-IFC
can be obtained from known outer bound regions for
the 2-IFC as follows.
By silencing one of the users, which is equivalent to give
the input signal of that user as side information
to the other two receivers, the channel effectively reduces to 
a 2-IFC to which known sum-rate outer bounds apply.
In particular, let:
\[
r_k=\log(1+|h_{kk}|^2), \ k\in\{1,2,3\},
\]
and let $r_{ij}$ be a sum-rate bound for a 2-IFC obtained by silencing
all but user $i$ and user $j$, $(i,j)\in\{1,2,3\}\times\{1,2,3\}$ with $i\not=j$.
Then, the sum-rate can be upper bounded by:
\begin{align}
r_{123}^{\rm(author)}
=\min\Big\{
&r_1+r_2+r_3, \ \ 
r_1+r_{23},
r_2+r_{13},
r_3+r_{12},
\nonumber\\&
\frac{r_{12}+r_{13}+r_{23}}{2}
\Big\}, 
\label{eq:known sum rate from 2-IFC}
\end{align}
where
$\rm author = Kra$ indicates that the 2-user rate bounds are obtained
from~\cite[Th.1]{kramer:it2004}, i.e., for example:
\begin{align*}
&r_{12}^{\rm(Kra)} = \min\!\Big\{\!
  \log(1+|h_{1,1}|^2+|h_{1,2}|^2) \!+\! \left[\log\left(\frac{1+|h_{2,2}|^2}{1+|h_{1,2}|^2}\right)\right]^+,
\\&\quad \log(1+|h_{2,1}|^2+|h_{2,2}|^2) +\left[\log\left(\frac{1+|h_{1,1}|^2}{1+|h_{2,1}|^2}\right)\right]^+
\Big\},
\end{align*}
and $\rm author = ETW$ indicates that the 2-user sum-rate bounds are obtained
from~\cite[Th.1]{etkin_tse_hua:withinonebit:subIt06}, i.e., for example:
\begin{align*}
r_{12}^{\rm(ETW)} 
  &=\log\left(1+|h_{1,2}|^2+\frac{|h_{1,1}|^2}{1+|h_{2,1}|^2}\right)
\\&+\log\left(1+|h_{2,1}|^2+\frac{|h_{2,2}|^2}{1+|h_{1,2}|^2}\right).
\end{align*}
The bounds $r_{123}^{\rm(Kra)}$ and $r_{123}^{\rm(ETW)}$
will be compared to $r_{123}^{\rm(Th.\ref{th:out1})}$ and $r_{123}^{\rm(Th.\ref{th:out2})}$, the sum-rate from Th.\ref{th:out1} and Th.\ref{th:out2}, respectively.

\begin{remark}[Other known outer bounds for the 2-IFC and their generalization to the $K$-user case]
Outer bounds known in the litterature for the Gaussian 2-IFC,
besides those in~\cite{kramer:it2004,etkin_tse_hua:withinonebit:subIt06}, are~\cite[Th.2]{kramer:it2004}
(which is tighter than~\cite[Th.1]{etkin_tse_hua:withinonebit:subIt06} for some weak interference parameters)
and~\cite{biaonoisy2008:IT09,venunoisy2008:IT09,canadanoisy,EtkinISIT2009} (which is sum-rate optimal in very weak 
interference).
It is left for future work to compare the $r_{123}$ bounds computed
according to~\reff{eq:known sum rate from 2-IFC} 
from these works with our Th.\ref{th:out1} and Th.\ref{th:out2}.

It is also left for future work to generalize the 2-IFC bounds
in~\cite[Th.2]{kramer:it2004} and in~\cite{biaonoisy2008:IT09,venunoisy2008:IT09,canadanoisy,EtkinISIT2009}
to the case of more than two users.
\end{remark}

\begin{remark}[Known outer bounds for some special K-IFC, $K\geq 3$]
As mentioned in the introduction, Gaussian K-IFC with special
structure for the channel matrix $\Hm$ have been considered in
the literature. In particular:
\begin{enumerate}

\item
The sum-rate of the ``cyclic mixed strong-very strong'' 3-IFC~\cite{segzin_3ifc:arXiv:1010.4911}
and of the ``cyclic symmetric'' 3-IFC in strong interference~\cite{ZhouYu:cyclicsym:CISS10} is given by
$r_{123}^{\rm(Kra)}$ in~\reff{eq:known sum rate from 2-IFC}.

\item
The capacity of the ``cyclic symmetric'' K-IFC in weak interference in~\cite{ZhouYu:cyclicsym:CISS10}
does not coincide with $r_{123}^{\rm(ETW)}$ in~\reff{eq:known sum rate from 2-IFC}
but it is a special case of our Th.\ref{th:out2}.

%

\item
It is left for future work to evaluate the MAC-based outer bound
in~\cite{josesriram:kifcmacbound:itw10} for the case of $K=3$ users and
compare it with our Th.\ref{th:out1} and Th.\ref{th:out2}.

\item
In our numerical examples we will also show the MAC sum-rate bound $r_{123}^{\rm(MAC)}$
obtained by letting all receivers cooperate so as to form
a MAC channel with three single-antenna transmitters
and a three-antenna receiver. The sum-capacity of this MAC channel is:
\begin{align}
r_{123}^{\rm(MAC)}
=\min_{\Sigmam_{123}} \log\left(\Id+\Hm \Hm^H \Sigmam_{123}^{-1} \right),
\label{eq:r123 MAC}
\end{align}
where the minimization is over all positive-definite noise covariance
matrix $\Sigmam_{123}$ constrained to have unit diagonal
elements (i.e., ``same conditional marginal'' constraint, see~\reff{eq:noise cov}).


\end{enumerate}
\end{remark}

\subsection{Evaluation of Th.\ref{th:out1} and Th.\ref{th:out2}}~\label{sec:gauss th1 and th2}

\begin{theorem}\label{th:out1 gauss}
For the Gaussian channel
Th.\ref{th:out1} reduces to:
\begin{align}
&R_1+R_2+R_3
\nonumber
\\&\leq I( Y_{1};X_{1}) + \min_{\rho: |\rho|\leq 1}       
   \Big\{I( Y_{1},Y_{2};X_{2}|X_{1})  
\nonumber
\\&  +\max\big\{
  I( Y_{1},Y_{2};X_{3}|X_{1},X_{2}), I( Y_{3};X_{3}|X_{1},X_{2})
\big\}   
    \Big\},
\label{eq:th:out1 gauss}
\end{align}
evaluated for iid $\Nc(0,1)$ inputs.
\end{theorem}
\begin{IEEEproof}
Since every mutual information term in
Th.\ref{th:out1} contains {\em all} inputs,
the ``Gaussian maximizes entropy'' principle assures that
jointly Gaussian inputs are optimal. 
Given the unitary power constraints, 
it is thus optimal to consider iid $\Nc(0,1)$ inputs
in~\reff{eq:th:out1}. 
Next we minimize the sum-rate with Gaussian inputs 
with respect to the noise covariance matrix:
\begin{align}
\Sigmam_{123}=
\begin{pmatrix}
1 & \rho & \rho_1 \\
\rho^* & 1 & \rho_2 \\
\rho_1^* & \rho_2^* & 1 \\
\end{pmatrix}
\defeq
\begin{pmatrix}
\Sigmam_{12} & \rhov \\
\rhov^H     & 1 \\
\end{pmatrix},
\label{eq:noise cov}
\end{align}
where $\rhov = (\rho_1,\rho_2)^T$ and $\Sigmam_{12}$ is the upper-left
$2\times2$ principal submatrix of $\Sigmam_{123}$.

We star by rewriting the sum-rate bound in~\reff{eq:th:out1} as:
\begin{align}
  R_{1}+R_{2}+R_{3}
&\leq 
 I( Y_{1};X_{1})
+I( Y_{1},Y_{2};X_{2}|X_{1})
\nonumber\\&
+I( Y_{1},Y_{2},Y_{3};X_{3}|X_{1},X_{2})
\label{eq:th1 rewritten}.
\end{align}
By the non-negativity of mutual information, the last term in~\reff{eq:th1 rewritten}
can lower bounded by:
\begin{align}
  &I( Y_{1},Y_{2},Y_{3};X_{3}|X_{1},X_{2})
\nonumber
\\&\geq \max\{I( Y_{1},Y_{2};X_{3}|X_{1},X_{2}), I( Y_{3};X_{3}|X_{1},X_{2})\}
\label{eq:th1 LB}.
\end{align}
The lower bound in~\reff{eq:th1 LB} is tight
if we can show that 
$Y_3$ and $(Y_1,Y_2)$ are one a degraded version of the other
when conditioned on $(X_1,X_2)$. Toward this goal, 
we whiten the noise in $(Y_1,Y_2)|_{(X_1,X_2)}$ 
and then perform  maximal ratio combining
so as to obtain an equivalent output:
\[
Y_{\rm eq} =  
\sqrt{
\begin{pmatrix}
 h_{1,3}^* &  h_{2,3}^* \\
\end{pmatrix}
\Sigma_{12}^{-1}
\begin{pmatrix}
 h_{1,3} \\
 h_{2,3} \\
\end{pmatrix}
} X_3
+
Z_{\rm eq}, 
\ Z_{\rm eq} \sim \Nc(0,1).
\]
Thus, 
\begin{itemize}
\item CASE 1:
if the SNR of $Y_{\rm eq}$ is larger than the SNR of 
$Y_3|_{(X_1,X_2)} \sim h_{33} X_3 + Z_3$, that is, if:
\begin{align}
\begin{pmatrix}
 h_{1,3}^* &  h_{2,3}^* \\
\end{pmatrix}
\Sigma_{12}^{-1}
\begin{pmatrix}
 h_{1,3} \\
 h_{2,3} \\
\end{pmatrix}
\geq |h_{33}|^2,
\label{eq:snr cond for case 1 = max for y1y2}
\end{align}
then $Y_3|_{(X_1,X_2)}$ is a degraded version of $Y_{\rm eq}$ and 
\begin{align*}
  &I( Y_{1},Y_{2},Y_{3};X_{3}|X_{1},X_{2})
=I( Y_{\rm eq};X_{3})
\\&   =I( Y_{1},Y_{2};X_{3}|X_{1},X_{2}).
\end{align*}

In this case, in order to determine the sum-rate in~\reff{eq:th:out1 gauss} 
we must still solve:
\begin{align}
\min_{\rho: |\rho|\leq 1}\{ I( Y_{1},Y_{2};X_{2},X_{3}|X_{1})\},
\label{eq:th:out1 gauss case 1}
\end{align}
where the minimization  in~\reff{eq:th:out1 gauss case 1}
is subject to the constraint in~\reff{eq:snr cond for case 1 = max for y1y2}.
The optimal $\rho$ in~\reff{eq:th:out1 gauss case 1}
{\em without} considering the constraint
from~\reff{eq:snr cond for case 1 = max for y1y2} can be obtained by applying
Lemma~\ref{lem:usefullemma} in the Appendix with:
\begin{align*}
\cv_1^T = (h_{1,2} \ h_{1,3}), \
\cv_2^T = (h_{2,2} \ h_{2,3}),
\end{align*}
to obtain that the optimal unconstrained $\rho$ is $\rho^{(1)}$ with:
\begin{align}
\rho^{(1)} 
 &= (t - \sqrt{t^2-1})\eu^{\jj \angle{\cv_1^H \cv_2}},
\label{eq:th:out1 gauss case 1 rho unconstrained}
\\
t &\defeq \frac{(1+\|\cv_1\|^2)(1+\|\cv_2\|^2) - |\cv_1^H \cv_2|^2 -1}{2|\cv_1^H \cv_2|}
\geq 1. \nonumber
\end{align}
The correlation coefficient $\rho^{(1)}$ 
can be the optimal solution for $\rho$ in~\reff{eq:th:out1 gauss} 
under certain conditions that we will discuss later on.

\item CASE 2:
If the condition in~\reff{eq:snr cond for case 1 = max for y1y2} is not satisfied,
then $Y_{\rm eq}$ is a degraded version of $Y_3|_{(X_1,X_2)}$  and
\begin{align*}
I( Y_{1},Y_{2},Y_{3};X_{3}|X_{1},X_{2})
=I(Y_{3};X_{3}|X_{1},X_{2}).
\end{align*}
%
In this case, in order to determine the sum-rate in~\reff{eq:th:out1 gauss} 
we must still solve:
\begin{align}
\min_{\rho: |\rho|\leq 1}\{ I( Y_{1},Y_{2};X_{2}|X_{1})\},
\label{eq:th:out1 gauss case 2}
\end{align}
where the minimization is subject to the complement condition of~\reff{eq:snr cond for case 1 = max for y1y2}.
The optimal $\rho$ in~\reff{eq:th:out1 gauss case 2}
{\em without} considering the constraint from the complement condition
of~\reff{eq:snr cond for case 1 = max for y1y2} can be obtained as follows.
In $I( Y_{1},Y_{2};X_{2}|X_{1})$ the signal $X_3$ acts as noise,
hence, by rewriting $(Y_{1},Y_{2})$ as:
\begin{align*}
Y_1'=
\frac{Y_1}{\sqrt{1+|h_{1,3}|^2}} 
&= \frac{h_{1,2}}{\sqrt{1+|h_{1,3}|^2}} X_2 +
   \frac{h_{1,3} X_3 + Z_1}{\sqrt{1+|h_{1,3}|^2}},
\\
Y_2'=
\frac{Y_2}{\sqrt{1+|h_{2,3}|^2}} 
&= \frac{h_{2,2}}{\sqrt{1+|h_{2,3}|^2}} X_2 +
   \frac{h_{2,3} X_3 + Z_2}{\sqrt{1+|h_{2,3}|^2}},
\end{align*}
we see that the correlation coefficient among the equivalent noises
in $Y_1'$ and $Y_2'$ is:
\begin{align*}
\rho' \defeq
\frac{h_{1,3} h_{2,3}^* + \rho}{\sqrt{(1+|h_{1,3}|^2)(1+|h_{2,3}|^2)}}.
\end{align*}
If the SNR of $Y_1'$ is smaller than the SNR of $Y_2'$, i.e.,:
\begin{align}
\frac{|h_{1,2}|^2}{1+|h_{1,3}|^2}
\leq 
\frac{|h_{2,2}|^2}{1+|h_{2,3}|^2}
\label{eq:th:out1 gauss case 2a condition}
\end{align}
then $Y_1'$ can be made a degraded version of $Y_2'$ if:
\begin{align}
&\rho'
=  
\frac{h_{1,2}}{\sqrt{1+|h_{1,3}|^2}}
\frac{\sqrt{1+|h_{2,3}|^2}}{h_{2,2}}
\Longleftrightarrow
\nonumber\\
&\rho^{(2a)}
= \frac{h_{1,2}}{h_{2,2}}(1+|h_{2,3}|^2)-h_{1,3}h_{2,3}^*.
\label{eq:th:out1 gauss case 2a rho unconstrained}
\end{align}
%
If the condition in~\reff{eq:th:out1 gauss case 2a condition} is not satisfied, 
then $Y_2'$ can be made a degraded version of $Y_1'$ if:
\begin{align}
&\rho'
=  
\frac{h_{2,2}}{\sqrt{1+|h_{2,3}|^2}}
\frac{\sqrt{1+|h_{1,3}|^2}}{h_{1,2}}
\Longleftrightarrow
\nonumber\\
&\rho^{(2b)}
= \frac{h_{2,2}}{h_{1,2}}(1+|h_{1,3}|^2)-h_{1,3}h_{2,3}^*.
\label{eq:th:out1 gauss case 2b rho unconstrained}
\end{align}
The correlation coefficients $\rho^{(2a)}$ and $\rho^{(2b)}$ 
can be the optimal solution for $\rho$ in~\reff{eq:th:out1 gauss} 
under certain conditions that we will discuss next.


\end{itemize}

The optimization over $\rho$ in~\reff{eq:th:out1 gauss}
can be carried out in closed form as follows (see for example~\cite[Sec.
II.C]{poorminax}).  
If $\rho^{(1)}$ in~\reff{eq:th:out1 gauss case 1 rho unconstrained}
satisfies the condition in~\reff{eq:snr cond for case 1 = max for y1y2}, then
$\rho=\rho^{(1)}$ is optimal.
If $\rho^{(2a)}$ in~\reff{eq:th:out1 gauss case 2a rho unconstrained}
satisfies the complement of the condition in~\reff{eq:snr cond for case 1 = max for y1y2},
$|\rho^{(2a)}|\leq 1$, and the condition in~\reff{eq:th:out1 gauss case 2a condition},
then $\rho=\rho^{(2a)}$ is optimal and 
the sum-rate in~\reff{eq:th:out1 gauss} becomes:
\begin{align}
&R_1+R_2+R_3 \leq
 \log\left(1+\frac{|h_{1,1}|^2}{1+|h_{1,2}|^2+|h_{1,3}|^2}\right)
\nonumber \\&
+\log\left(1+\frac{|h_{2,2}|^2}{1+|h_{2,3}|^2}\right)
+\log\left(1+|h_{33}|^2\right).
\label{eq:th:out1 gauss case 2a}
\end{align}
If $\rho^{(2b)}$ in~\reff{eq:th:out1 gauss case 2b rho unconstrained}
satisfies the complement of the condition in~\reff{eq:snr cond for case 1 = max for y1y2}
$|\rho^{(2b)}|\leq 1$, and the complement of the condition in~\reff{eq:th:out1 gauss case 2a condition}, then
$\rho=\rho^{(2b)}$ is optimal and 
the sum-rate in~\reff{eq:th:out1 gauss} becomes:
\begin{align}
&R_1+R_2+R_3 \leq
\log\left(1+\frac{|h_{1,1}|^2+|h_{1,2}|^2}{1+|h_{1,3}|^2}\right)
\nonumber \\&
+\log\left(1+|h_{33}|^2\right).
\label{eq:th:out1 gauss case 2b}
\end{align}
In all other cases, the optimal $\rho$ in~\reff{eq:th:out1 gauss} 
is such that the condition in~\reff{eq:snr cond for case 1 = max for y1y2}
holds with equality, that is, $\rho = \rho^{(3)}$ is optimal
with:
\begin{align}
\left|
\rho^{(3)} -  \frac{h_{1,3}h_{2,3}^*}{|h_{3,3}|^3}
\right|
=
\sqrt{1-\frac{|h_{1,3}|^2}{|h_{3,3}|^3}}
\sqrt{1-\frac{|h_{2,3}|^2}{|h_{3,3}|^3}}.
\label{eq:th:out1 gauss case 3 rho unconstrained}
\end{align}
For cases~1 and~3 the closed-form expression for the sum-rate in~\reff{eq:th:out1 gauss}
is quite involved and we do not explicitly write it here for sake of space.
\end{IEEEproof}


\begin{theorem}\label{th:out2 gauss}
For the Gaussian channel Th.\ref{th:out2} reduces to:
\begin{align}
&R_1+R_2+R_3
\leq \min_{\pi} \Big\{ f_{k,\pi_k}\Big\},
\label{eq:th:out2 gauss}
\end{align}
where $\pi$ is a permutation of the vector $(1,2,3)$ and where
\begin{align*}
f_{j,k} = \log\left(\frac{1+|\rv_j \rv_{\backslash k}^H| (q+\sqrt{q^2-1})}{1+ \| \rv_{\backslash k}\|^2}\right),
\end{align*}
for $\rv_k = (h_{k,1},h_{k,2},h_{k,3})$
(i.e., the set of channel coefficients seen at receiver $k$
arranged in a row vector), with
$[\rv_{\backslash k}]_u = [\rv_k]_u \delta_{k-u}$
(i.e., $\rv_{\backslash k}$ equals $\rv_k$ except for the $k$-th entry which is zero),
and for $(j,k)\in\{1,2,3\}^2$
\begin{align*}
q = \frac{(1+ \| \rv_j\|^2)(1+ \| \rv_{\backslash k}\|^2)-|\rv_j \rv_{\backslash k}^H|^2 -1}{2 |\rv_j \rv_{\backslash k}^H|}\geq 1.
\end{align*}
\end{theorem}
\begin{IEEEproof}
In the Gaussian case, for $k\in\{1,2,3\}$, we have:
\begin{align*}
&Y_{\backslash k}
  \sim Y_k-h_{k,k}X_k
\\&= \sum_{u\in\{1,2,3\}\backslash \{k\}}h_{k,u}X_u + Z_{\backslash k}
= \rv_{\backslash k}  (X_1, X_2, X_3)^T+ Z_{\backslash k},
\end{align*}
where $Z_{\backslash k}\sim\Nc(0,1)$ whose correlation with the Gaussian 
noise on the channel output can be chosen so as to tighten the sum-rate
bound in~\reff{eq:th:out2}. Next, by applying Lemma~\ref{lem:usefullemma}
in the Appendix, we obtain: 
\begin{align*}
&\min_{\rho=\E[Z_j Z_{\backslash k}^*] : |\rho|\leq 1}
h(Y_j|Y_{\backslash k}) - h(Z_{\backslash k}|Z_j)
\\&=\min_{\rho=\E[Z_j Z_{\backslash k}^*] : |\rho|\leq 1}
I(Y_j; X_1, X_2, X_3|Y_{\backslash k}) = f_{j,k},
\end{align*}
for $f_{j,k}$ defined above.
By considering all the permutations in~\reff{eq:th:out2} such that the
side information set $\{S_k, k\in\{1,2,3\} \}$ coincides with
the set $\{Y_{\backslash k}, k\in\{1,2,3\} \}$ we finally get
the claimed expression for the sum-rate in~\reff{eq:th:out2 gauss}.
\end{IEEEproof}

\subsection{Numerical Example}~\label{sec:gauss examples}

\begin{figure*}
\centerline{
\subfigure[]{\includegraphics[width=3.5in]{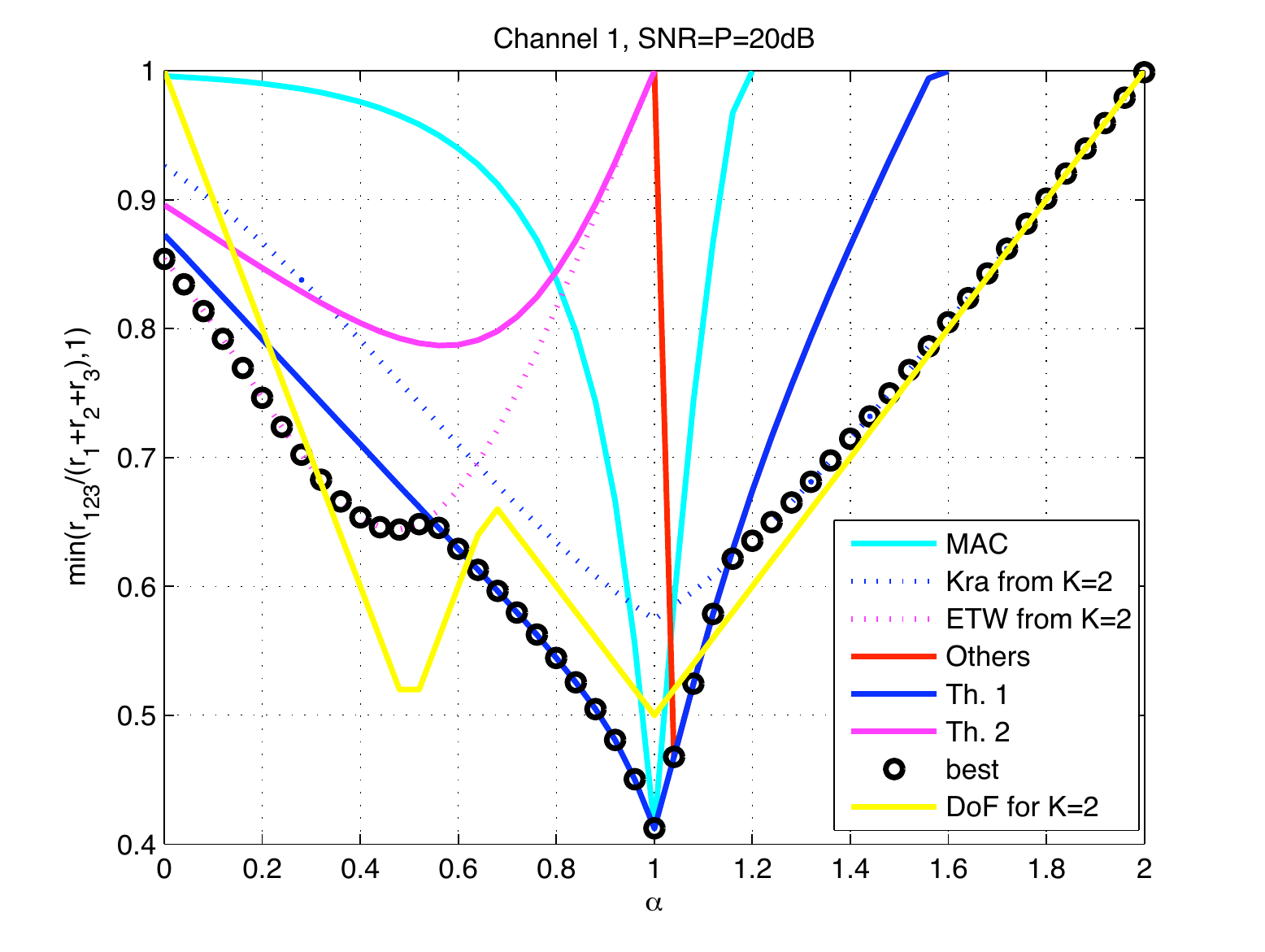}
\label{fig11}} \hfil
\subfigure[]{\includegraphics[width=3.5in]{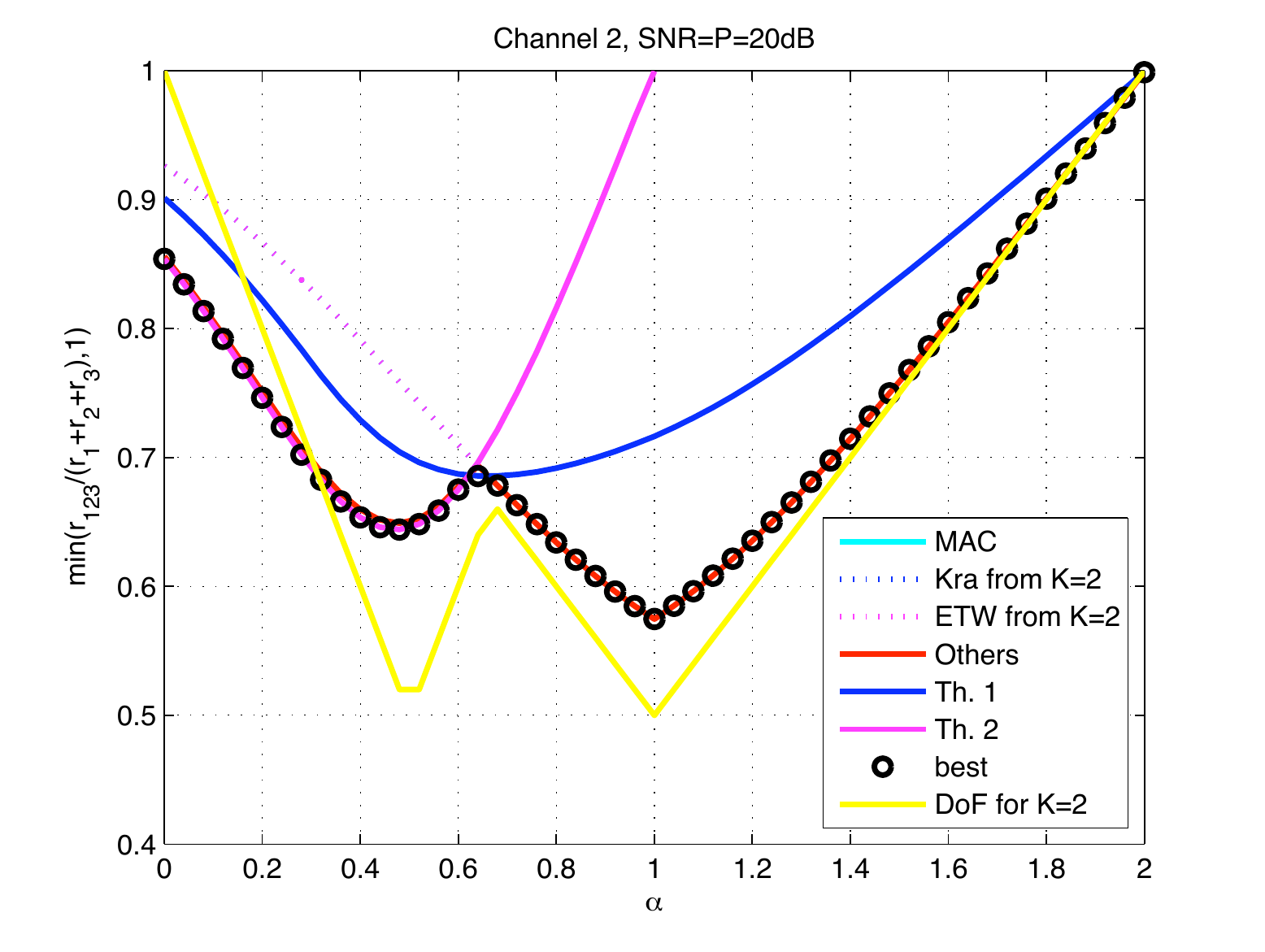}
\label{fig12}}  \hfil
} 
\caption{
${\rm DoF}(P) \defeq\min\{r_{123}/(r_1+r_2+r_3),1\}$ vs. $\alpha = \log(P|h|^2)/\log(P)$
at $P=20$dB for
(a) the ``fully symmetric'' channel, and
(b) the ``cyclic symmetric'' channel.
}
\label{fig2}
\end{figure*}

In this section we compare our proposed sum-rate outer bound
in Th.\ref{th:out1 gauss} and in Th.\ref{th:out2 gauss}
with $r_{123}^{\rm(Kra)}$, $r_{123}^{\rm(ETW)}$ and $r_{123}^{\rm(MAC)}$
as defined in subsection~\ref{sec:gauss known sumrate}
for the following class of channels:
\[
\Hm = 
\begin{pmatrix}
 1 & h_1 & h_2\\
 h_2 & 1 & h_1\\
 h_1 & h_2 & 1\\
\end{pmatrix}\sqrt{P}
\]
The ``fully symmetric'' channel 
is obtained for
$h_1
=h_2=h,$ 
the ``cyclic symmetric'' channel of~\cite{ZhouYu:cyclicsym:CISS10} for
$h_1=h, \ 
 h_2=0,$
and the ``mixed strong-very strong'' channel of~\cite{segzin_3ifc:arXiv:1010.4911} for
$|h_1|^2 \geq 1, \
 |h_2|^2 \geq {1+|h_1|^2+1/P}.$
In Fig.~\ref{fig2} we plot the function: 
\[
{\rm DoF}(P) \defeq
\min\{r_{123}/(r_1+r_2+r_3),1\}
\] 
vs. $\alpha = \log(|h|^2 P)/\log(P)$ for $P=20$dB.
By the definition in~\reff{eq:known sum rate from 2-IFC},
$r_{123}^{\rm(Kra)}/(r_1+r_2+r_3) \leq 1$ as well as
$r_{123}^{\rm(ETW)}/(r_1+r_2+r_3) \leq 1$; however
$r_{123}^{\rm(Th.1)}$ in~\reff{eq:th:out1 gauss},
$r_{123}^{\rm(Th.2)}$ in~\reff{eq:th:out2 gauss} and
$r_{123}^{\rm(MAC)}$ in~\reff{eq:r123 MAC}
need not be smaller than the sum of the single-user 
rate bounds. The function ${\rm DoF}(P)$
can be thought of as the DoF of the channel
at a finite $\snr=P$; the DoF of the channel normalized
by the number of users, as per definition
commonly accepted in the literature,
is ${\rm DoF}=\lim_{P\to\infty}{\rm DoF}(P)$.

The ${\rm DoF}(P)$ of the ``fully symmetric'' channel is presented in Fig.~\ref{fig11}.
From~\cite{jafar:dofKifc:it10}, the ${\rm DoF}=\lim_{P\to\infty}{\rm DoF}(P)$
is as for the 2-user channel~\cite{etkin_tse_hua:withinonebit:subIt06} for any $K\geq 2$
except at $\alpha=1$ where it equals $1/K$ (because
the channel matrix is always full rank for any $h\not=1$
but is has unit rank for  $h=1$).  From  Fig.~\ref{fig11}
we see that the DoF (yellow line) does not accurately predict
performance at $P=20$dB yet.  For $h=1$, that is when all
channel outputs are statistically equivalent, the crude MAC
bound (cyan line) is optimal.  The new bound $r_{123}^{\rm(Th.1)}$
(solid blue line)
is the tightest for approximately $\alpha\in[0.54, 1.15]$ (the 
range enlarges for smaller $P$).  For approximately $\alpha\leq 0.54$
the bound $r_{123}^{\rm(ETW)}$ (dotted magenta line) is the tightest and for 
approximately $\alpha\geq 1.15$ the bound 
$r_{123}^{\rm(Kra)}$ (dotted blue line) is the tightest.
The line labeled with ``Others'' (red line) refers to the performance
of the genie aided bound in~\cite{cadambe:notseparable:it09}, which
is valid for $\alpha\geq 1$ only
and coincides with $r_{123}^{\rm(Th.1)}$.

The ${\rm DoF}(P)$ of the ``cyclic  symmetric'' channel is presented in Fig.~\ref{fig12}.
From~\cite{ZhouYu:cyclicsym:CISS10}, ${\rm DoF}=\lim_{P\to\infty}{\rm DoF}(P)$
is as for the 2-user channel~\cite{etkin_tse_hua:withinonebit:subIt06} (in this case
there are no discontinuity since the channel matrix is always full rank).
From  Fig.~\ref{fig12}
we see that the DoF (yellow line) does not accurately predict
performance at $P=20$dB yet. 
In this case, the bound $r_{123}^{\rm(Th.2)}$ is the tightest
for $\alpha\leq 2/3$ and $r_{123}^{\rm(Kra)}=r_{123}^{\rm(ETW)}$
(dotted magenta and blue lines) is the tightest and for 
$\alpha\geq 2/3$.  The line labeled with ``Others'' (red line)
refers to the outer bound in~\cite{ZhouYu:cyclicsym:CISS10}; the bound
$r_{123}^{\rm(Th.2)}$ is tighter than~\cite{ZhouYu:cyclicsym:CISS10} because
of the optimized correlation coefficient between the noise
of the side information signal and the noise of the channel output
used in $r_{123}^{\rm(Th.2)}$, as opposed to independent noises
as used in~\cite{ZhouYu:cyclicsym:CISS10};
as remarked in~\cite{EtkinISIT2009}, at high $\snr$ the optimal
correlation is zero.

The performance of the ``mixed strong-very strong'' channel
is not shown here because we know from~\cite{segzin_3ifc:arXiv:1010.4911}
that $r_{123}^{\rm(Kra)}$ is optimal.

\section{Conclusions}\label{sec:conc}
In this work we developed a framework to derive sum-rate
upper bounds for the general memoryless interference channel
with three source-destination pairs.  For the Gaussian channel,
the proposed bound is the tightest known for certain channel parameters. 

\section*{Acknowledgment}
This work was partially funded by NSF under award number 0643954.
The contents of this article are solely the responsibility of the authors and
do not necessarily represent the official views of the NSF.

\appendix

\begin{lemma}\label{lem:usefullemma}
For two MISO AWGN channels
\begin{align*}
Y_c = \cv_c^H \Xv + Z_c, \ c\in\{1,2\},
\end{align*}
where $\Xv\sim \Nc(0, \Id)$ is independent of $Z_c\sim\Nc(0,1)$, $c\in\{1,2\}$,
and $\cv$ is a column vector of the same dimension of the input $\Xv$,
we have
\begin{align*}
  &\min_{\rho \defeq \E[Z_1 Z_2^*]: \ |\rho|\leq 1} \Big\{I(Y_1; \Xv|Y_2) \Big\}
\\&= \log\left(1+ |\cv_1^H \cv_2| (t+\sqrt{t^2-1})\right)
-\log(1+\|\cv_2\|^2),
\end{align*}
with
\begin{align*}
t \defeq \frac{(1+\|\cv_1\|^2)(1+\|\cv_2\|^2) - |\cv_1^H \cv_2|^2 -1}{2|\cv_1^H \cv_2|}
\geq 1,
\end{align*}
and where the  minimum is attained by:
\begin{align*}
 \rho^{\rm(opt)} \defeq (t - \sqrt{t^2-1})\eu^{\jj \angle{\cv_1^H \cv_2}} \in[0,1] \quad \forall t\geq 1.
\end{align*}
\end{lemma}
\begin{IEEEproof}
Assume $\cv_1^H \cv_2\not = 0$. We have:
\begin{align*}
&\eu^{I(Y_1; \Xv|Y_2)}
  =    \frac{1+\|\cv_1\|^2 - \frac{|\cv_1^H \cv_2 + \rho|^2}{1+\|\cv_2\|^2}}{1-|\rho|^2}
\\&\geq \frac{(1+\|\cv_1\|^2)(1+\|\cv_2\|^2) - (|\cv_1^H \cv_2| + |\rho|)^2}{1-|\rho|^2}\frac{1}{1+\|\cv_2\|^2}
\\&\geq \left(1+2|\cv_1^H \cv_2| \min_{\rho: |\rho|\leq 1} \frac{t - |\rho| }{1-|\rho|^2}\right)\frac{1}{1+\|\cv_2\|^2}
\\&=    \left(1+ \frac{|\cv_1^H \cv_2|}{t-\sqrt{t^2-1}}\right)\frac{1}{1+\|\cv_2\|^2}
\\&=    \left(1+ |\cv_1^H \cv_2| (t+\sqrt{t^2-1})\right)\frac{1}{1+\|\cv_2\|^2},
\end{align*}
as claimed.
\end{IEEEproof}

\begin{remark}
The function $\min_{\rho}\{ I(Y_1; \Xv|Y_2)\}$ is decreasing in the 
angle between the vectors $\cv_1$ and $\cv_2$.
For $|\cv_1^H \cv_2 |^2 = \|\cv_1\|^2 \|\cv_2\|^2$ (i.e., parallel channel vectors),
we have:
\begin{align*}
  \min_{\rho} I(Y_1; \Xv|Y_2)
= \log\left(\frac{1+\max\{\|\cv_1\|^2,\|\cv_2\|^2\}}{1+\|\cv_2\|^2}\right)
\end{align*}
as for SISO (degarded) channels.
For $|\cv_1^H \cv_2| = 0$ (i.e., orthogonal channel vectors), one can easily see that $|\rho|=0$ is optimal and thus:
\begin{align*}
   \min_{\rho} I(Y_1; \Xv|Y_2)
  =\log(1+\|\cv_1\|^2).
\end{align*}
\end{remark}

\begin{remark}
By using $\rho=0$ we would get:
\begin{align*}
 &I(Y_1; \Xv|Y_2)|_{\rho=0}
  =\log\left(1+\|\rv_1\|^2 - \frac{|\cv_1^H \cv_2|^2}{1+\|\rv_2\|^2}\right)
\\&\in
\left[
 \log\left(1+ \frac{\|\rv_1\|^2}{1+\|\rv_2\|^2}\right), \
 \log\left(1+ \|\rv_1\|^2 \right)
\right].
\end{align*}
\end{remark}


\bibliographystyle{IEEEtran}
\bibliography{career_biblio_v3}

\end{document}